\documentclass[a4paper,pra,twocolumn,superscriptaddress]{revtex4-1}
\usepackage[english]{babel}
\usepackage[utf8]{inputenc}
\usepackage{graphicx,epsfig,xcolor}
\usepackage{latexsym}
\usepackage{amsfonts}
\usepackage{amsmath}
\usepackage{hyperref}
\usepackage{placeins}
\onecolumngrid
\usepackage{polski}
\newcommand{\Tr}{\mathrm{Tr}}
\def\ket|#1>{| #1 \rangle}
\def\bra<#1|{\langle #1 |}
\def\<{\langle}
\def\>{\rangle}
\def\{{\lbrace}
\def\}{\rbrace}
\def\({\left(}
\def\){\right)}
\def\beq{\begin{equation}}
\def\eeq{\end{equation}}

\begin{document}

\title{Genuine multipartite entanglement in a one-dimensional Bose-Hubbard model\\  with frustrated hopping}

\author{Sudipto Singha Roy}
\affiliation{Pitaevskii BEC Center, CNR-INO and Dipartimento di Fisica, Università di Trento, I-38123 Trento, Italy}
\affiliation{INFN-TIFPA, Trento Institute for Fundamental Physics and Applications, Trento, Italy}

\author{Leon Carl}
\affiliation{Pitaevskii BEC Center, CNR-INO and Dipartimento di Fisica, Università di Trento, I-38123 Trento, Italy}

\author{Philipp Hauke}
\affiliation{Pitaevskii BEC Center, CNR-INO and Dipartimento di Fisica, Università di Trento, I-38123 Trento, Italy}
\affiliation{INFN-TIFPA, Trento Institute for Fundamental Physics and Applications, Trento, Italy}
\begin{abstract}Frustration and quantum entanglement are two exotic quantum properties in quantum many-body systems. However, despite several efforts, an exact relation between them remains elusive. In this work, we explore the relationship between frustration and quantum entanglement in a physical model describing strongly correlated ultracold bosonic atoms in optical lattices. In particular,   we consider the one-dimensional Bose--Hubbard model comprising both nearest-neighbor ($t_{1}$) and frustrated next-nearest neighbor ($t_{2}$) hoppings and examine how the interplay of onsite interaction ($U$) and hoppings results in different quantum correlations dominating  in the ground state of the system.  We then analyze the behavior of quantum entanglement in the model. In particular, we compute genuine multipartite entanglement as quantified through the generalized geometric measure and make a comparative study with bipartite entanglement and other relevant order parameters.   We observe that genuine multipartite entanglement has a very rich behavior throughout the considered parameter regime and frustration does not necessarily favor generating a high amount of it. Moreover, we show  that in the region with strong quantum fluctuations, the particles remain highly delocalized in all momentum modes and share a very low amount of both bipartite and multipartite entanglement. Our work illustrates the necessity to give separate attention to  dominating ordering behavior and quantum entanglement in the ground state of strongly correlated systems.
\end{abstract}

\date{\today}

\maketitle
\section{Introduction}
\begin{figure*} 
  \includegraphics[width=11cm]{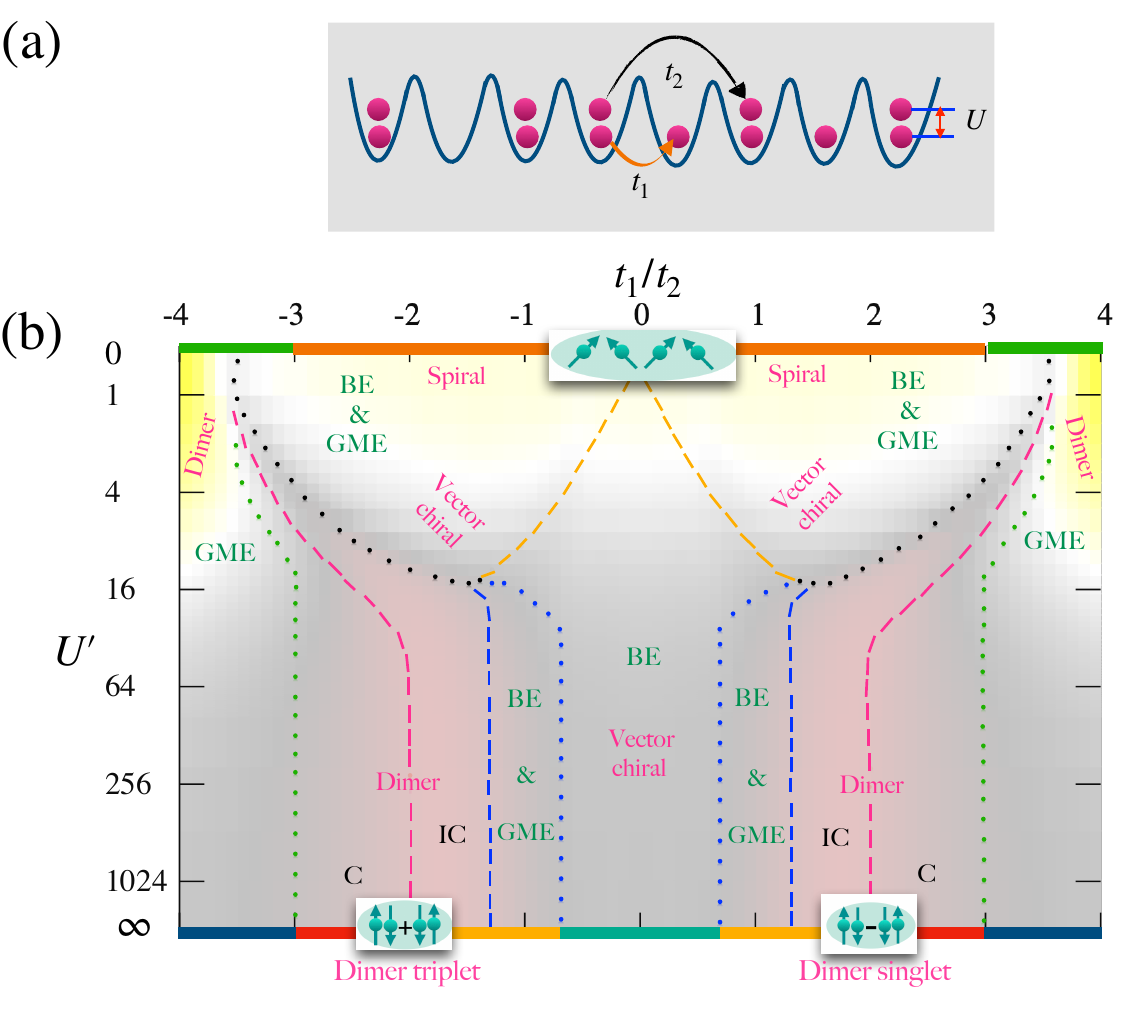}
\caption{ (a) Schematic of arrangement of quantum particles in a one-dimensional (1D) lattice  described by the Bose--Hubbard model comprising   NN ($t_1$) and NNN  ($t_2$) hoppings and onsite interaction $U$. Competition between those parameters leads to frustration and strong quantum fluctuations. (b) We schematically show the  distribution of different dominating quantum correlations in the GS of the above  Bose--Hubbard model, in the $U'-t_1/t_2$ plane. Here, using the analysis performed in Sec.  {\!} \ref{chiral_order} and \ref{dimer_order}, we  mark the regions with different dominating  orders and use dotted lines to  approximately separate them. The subsequent analysis of quantum entanglement presented in Sec. {\!} \ref{sec:entanglement} helps us in identifying the regions with dominating genuine multipartite entanglement (GME) and bipartite entanglement (BE), quantified through generalized geometric measure and half-chain entanglement entropy, respectively. For small $U$, the dashed yellow and dotted black lines enclose the region with both high  BE and GME. Similar region for high $U$ is  marked by the dashed and dotted blue lines. Additionally, we draw the approximate boundary (dashed magenta lines) separating predominating commensurate (C)  and incommensurate (IC) phase orderings as discussed in Sec. {\!}\ref{sec:population}.
}
  \label{fig:schematic1}    
\end{figure*}

The past few decades have witnessed many
 interesting theoretical developments in the field of quantum
 information theory. On one hand, the laws of quantum mechanics 
have been exploited to propose quantum computation and quantum information schemes 
that often surpass their classical counterparts
 \cite{dense_coding,teleportation,shor_algo,cryptography,Nielsen_chuang,
 Wilde,ent,quantum_inf1}. On the other hand, tools from quantum information theory
 have been used to unveil many interesting phenomena of physical 
systems that belong to a wide variety of interdisciplinary fields, 
ranging from condensed matter systems 
\cite{ent_many_body1,ent_many_body2,ent_many_body3,ent_many_body4,Laflorencie} 
over high-energy physics 
\cite{Dalmonte,ent_high_eng1,ent_high_eng2,ent_high_eng3} to holography 
\cite{ent_holography6,ent_holography1,ent_holography2,ent_holography3,ent_holography4,
ent_holography5}, etc.   
 In recent years,
 promising developments have been reported in designing
state-of-the-art quantum technologies using quantum many-body
 systems that include trapped ions \cite{Trapped_ion_Qinfo1,Trapped_ion_Qinfo2,Trapped_ion_Qinfo3,
 Trapped_ion_Qinfo4}, superconducting quantum circuits \cite{SC_QIP1,SC_QIP2,SC_QIP3,SC_QIP4}, 
 silicon-based devices \cite{Silicon_Qinfo1,Silicon_Qinfo2,Silicon_Qinfo3,Silicon_Qinfo4}, photonic systems \cite{photonic_Qinfo1,photonic_Qinfo2,photonic_Qinfo3,photonic_Qinfo4,
 photonic_Qinfo5,photonic_Qinfo6}, and atomic systems \cite{ultracold_ref1,ultracold_ref2,ultracold_ref3},   which can efficiently produce large amounts of entanglement \cite{bloch,atom_ent, SC_ent,Friis,Jurcevic}. 
As in many cases quantum entanglement remains the key resource of quantum technologies, a primary step in the assessment of a quantum device demands complete characterization of its entanglement properties. Besides this, in the literature, there have been a plethora of works  \cite{ent_many_body1,ent_many_body2,ent_many_body3,Guhne,
cirac1,cirac2,Sarandy,Laflorencie,Heyl,Giampolo,hsd,chiara1} where alongside conventional order parameters, quantum entanglement has been considered as an efficient detector of quantum phase boundaries in exotic quantum many-body systems.


Quantum entanglement shared between a large number of parties often gives rise to a highly intricate   form of  quantum correlations, namely, multipartite entanglement (ME) \cite{chiara2,ggm1, ggm2,ggm3,ggm4,ggm5,ggm6,chiara2,Heyl,hsd,ssr0,ssr1,ssr2,ssr3}. It is known that ME can serve as a resource in the implementation of novel quantum schemes such as measurement-based quantum computation \cite{MBQC}, quantum cryptography \cite{ME_cryptography}, quantum sensing \cite{ME_sensing1,ME_sensing2}, quantum error correction \cite{ME_Error_corr}, etc. Moreover, there are instances where ME performs as a better identifier of quantum phase boundaries than bipartite entanglement (BE) \cite{Giampaolo,haldar,oliveira}. 
 However, quantification of ME in complex quantum many-body systems is an extremely  challenging task.  Unlike BE, a computable measure of ME is difficult to construct even for pure quantum states.  In particular, the characterization of genuine multipartite entanglement (GME) requires full knowledge of the  entanglement distribution in all possible bipartitions of the system \cite{ggm1, ggm2,ggm3,ggm4,chiara2,Heyl,GME_concurrence}. 
 Hence, a complete characterization of ME even for a finite-size system is an important albeit difficult task.


In this work, we consider one paradigmatic model of strongly correlated quantum particles on an optical lattice, namely the Bose--Hubbard model \cite{ref_BH1,ref_BH2,Eckardt1} with frustration which is introduced by the inclusion of beyond nearest-neighbor hopping (Fig. {\!}\ref{fig:schematic1}(a)), and characterize its bipartite and multipartite entanglement  properties. The limiting cases of the model have been well explored in previous works. For instance, in the hard-core boson limit in the ground state (GS) configuration, there exists a competition between vector chiral and several dimer orders resulting from strong quantum fluctuations \cite{Majumder_ghosh, sato-et-al1,sato-et-al2,Schmied}. In contrast,   near the limit of vanishing onsite interaction, the quantum phase reminiscent to the classical spin spiral phase remains a dominating feature of the GS. In our work, we aim to explore how the  GS characteristics change as a result of the interplay of finite onsite interaction interpolating between the two limiting cases and frustration in the system. In particular, we aim at identifying regions in the parameter space comprising different ordering tendencies  and make a comparative study with the entanglement properties. As a measure of GME, we consider the generalized geometric measure (GGM) \cite{ggm1, ggm2,ggm3,ggm4,chiara2} and observe its rich behavior in the considered parameter regime.  For a wide region in the parameter space, the behavior of both BE and GME  remain compatible with that of chiral order in that region. However, unlike BE, when NNN hopping dominates ($|t_1|/t_2<1$), GME becomes very low.  In contrast, in the region where strong quantum frustration drives the system to assume dimer order, we observe a significantly lower value of both BE and GME. Based on our analysis, we argue that the behavior obtained for the hard-core boson limit of the model approximately translates up to a finite but moderate value of onsite interaction.  Moreover, our work suggests  entanglement properties of quantum many-body  systems do not always  manifest in the behavior of different ordering tendencies  and deserve separate analysis.

The article is organized as follows. In Sec. {\!}\ref{sec:model}, we introduce the model Hamiltonian that we consider in our work and discuss its two limiting cases. In Sec. {\!}\ref{sec:results}, we discuss the behavior of different order parameters obtained for the GS of the model, leading to the sketch  displayed in Fig. {\!}\ref{fig:schematic1}(b). Thereafter, in Sec. {\!}\ref{sec:entanglement}, we discuss the entanglement properties of the model and compare them with the results obtained in Sec. {\!}\ref{sec:results}. We finally conclude and discuss our future plans in Sec. {\!}\ref{conclusion}.

\begin{figure*}
  \includegraphics[width=17cm]{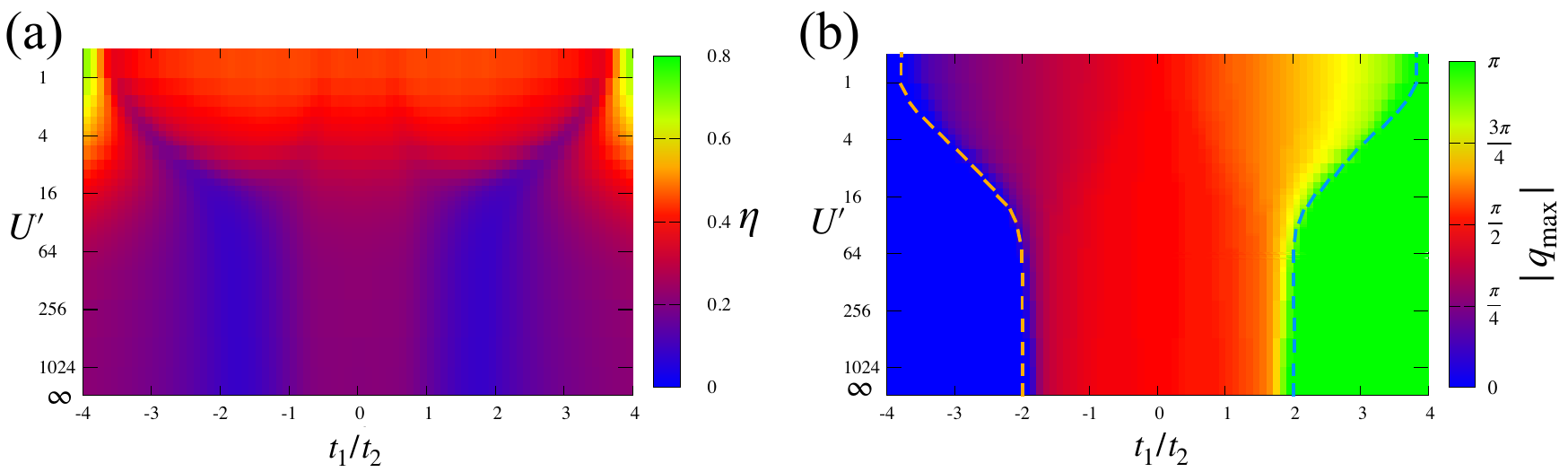}
  \caption{Variation of momentum density in  the GS of the Bose--Hubbard model   with the NN and NNN hopping ratio  and contact interaction.  (a) Behavior of maximum occupation density  $\eta =\max_q \langle {\hat{b}_q}^\dagger \hat{b}_q \rangle/N$ obtained for OBC. For moderately large values of $U$, $\eta$ takes its minimal value in the region $1<|t_1|/t_2<2$. This suggests the bosons are delocalized or  maximally spread   over all momentum modes, indicating  regions with strong quantum fluctuations. For small $U$ and PBC, bosons condense into a single mode (collinear phases at $|t_1|/t_2>4$)
   or two modes (spiral phase). We denote the maximally populated momentum mode by $q_{\max}$ and plot it in panel (b). At $U\rightarrow 0$,  it minimizes the dispersion relation given in Eq. {\!}(\ref{eqn:dispersion}).   The dashed yellow (blue) line (Lifshitz line) distinguishes the regions with predominating commensurate   phase order observed for $q_{\max}=0$ ($|q_{\max}|=\pi$) and the noncommensurate  phase order observed for $0<|q_{\max}|<\frac{\pi}{2}$ ($\frac{\pi}{2}<|q_{\max}|<\pi$).  Data for  $N=10$ and $M=20$. }
  \label{fig:density}    
\end{figure*}

\section{Model}
\label{sec:model}
In this section, we introduce the model Hamiltonian that we consider in our work, which is the  Bose--Hubbard (BH) model in 1D  with both nearest-neighbor (NN)  and next-nearest-neighbor (NNN) tunneling terms, given by 
\begin{eqnarray}
\hat{H}&=&-t_1\sum_{\langle i,j\rangle}({\hat{b}_i}^\dagger\hat{b}_{j}+{\hat{b}_j}^\dagger\hat{b}_{i})-t_2 \sum_{\langle \langle i,j\rangle \rangle } ({\hat{b}_i}^\dagger\hat{b}_{j}+{\hat{b}_{j}^\dagger}\hat{b}_{i})\nonumber\\&+&\frac{U}{2}\sum_i \hat{n}_i\left(\hat{n}_i-1\right),
\label{eqn_Ham}
\end{eqnarray}
where $\hat{b}_i$  is the bosonic annihilation operator at site $i$, $t_1$ ($t_2$) corresponds to the NN (NNN) hopping amplitude,  $U$  is the on-site interaction energy, and $n_i$ is the number of bosons at site $i$. The total number of sites is given by $M$ and $N=\sum_in_i$ denotes the number of particles.  Physically, such a model may be realized in an optical lattice in a zig-zag ladder configuration \cite{Eckardt1,Cabedo}, where $t_1/t_2$ can be tuned via the ladder width. The sign of $t_1/t_2$ can be adjusted through a synthetic magnetic field \cite{linger, weiss,struck,Eckardt2,HaukeQHE,Dalibard}.

Before going into details of our analysis, we review two limiting cases of the model. 
\subsection{Free bosons, $U=0$ limit }
\label{free_bosons}
For the $U=0$ limit, and  periodic boundary conditions (PBC),  the following relation 
\begin{equation}
\hat{b}_{j}=\frac{1}{\sqrt{M}}\sum_{q}e^{i jq}\hat{b}_q, 
\end{equation}
diagonalizes the above Hamiltonian to 
\begin{eqnarray}
\hat{H}&=&\sum_q \mathcal{E}_q {\hat{b}_q}^\dagger \hat{b}_q.
\end{eqnarray}
The dispersion relation in this free-boson limit is  given by 
\begin{eqnarray}
\mathcal{E}_q =-2t_1\cos q-2t_2\cos 2q.
\label{eqn:dispersion}
\end{eqnarray}
Depending on the  sign of  $t_2$, two distinct regimes appear:
\begin{itemize}
\item[] i) For  $t_2>0$: $\mathcal{E}_q$ is minimized by having all atoms in the $q=0$ mode, for $t_1>0$ ($q=\pi$ mode, for $t_1<0$).\\
\item[] ii)  For $t_2<0$: There is a competition between the first and second term in $\mathcal{E}_q$, which independently would be minimized by $q=0$ for $t_1>0$ ($q=\pi$, for $t_1<0$) and $q=\pm \pi/2$, respectively. 
As a result, the maximally populated $q$ mode that yields minimum  $\mathcal{E}_q$ becomes
\begin{equation}
q_{\max}=\begin{cases}
0, \text{for}\hspace{.1cm} t_1/t_2\leq -4,\\
\pm \cos^{-1}(-\frac{t_1}{4t_2}),\hspace{.1cm}  \text{for}\hspace{.1cm} -4< t_1/t_2<4, \\
\pi, \text{for}\hspace{.1cm}  t_1/t_2\geq4.
\end{cases}
\label{eqn:t1-t2-range}
\end{equation}
\end{itemize}
Hence, $q_{\max}$ behaves exactly as the pitch angle of the helical spin arrangements of the corresponding frustrated classical spin model.

In our analysis,  we consider the  second scenario, and the primary focus will be in the region $-4\leq t_1/t_2\leq 4$, that consists of a ferromagnetic phase for $t_1/t_2<-4$, a classical spiral phase for $-4<t_1/t_2<4$, and an antiferromagnetic phase for $t_1/t_2>4$. To mitigate the effect of incommensurate pitch angles in our finite-size numerics, in the remainder of this work we  consider open boundary conditions (OBC).   One should note that when we consider $U\neq0$ or OBC, the above dispersion relation will not reflect the actual GS population and we expect the particles to occupy other momentum modes as well.

\begin{figure*}
  \includegraphics[width=17cm]{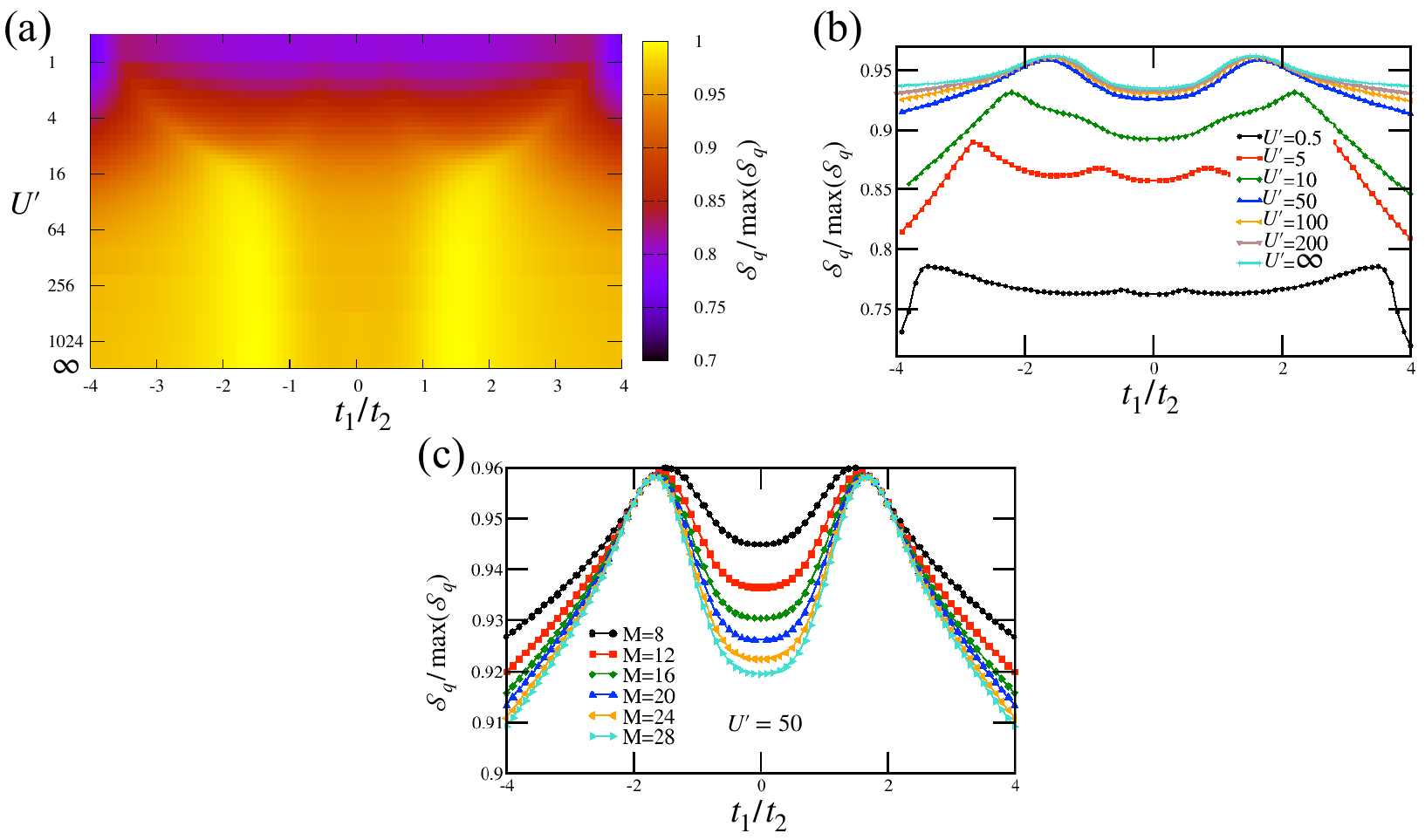}
 \caption{Spread of population of quantum particles across different momentum modes.  (a) Entropy of  momentum population  in GS, as quantified by $\mathcal{S}_q$ for  $N=10$, $M=20$. In the plot, we scaled $\mathcal{S}_q$ by its maximum value $\max(\mathcal{S}_q)=\ln N_q$.   (b) Cuts of $\mathcal{S}_q$ for fixed $U$, for $N=10$, and $M=20$. In panel (c), we plot the scaling of $\mathcal{S}_q/\max{(\mathcal{S}_q)}$ with different system sizes, $M=8, 12, 16, 20, 24$, and 28 for $U'=50$. The scaling suggests even at moderate system size $\mathcal{S}_q/\max{(\mathcal{S}_q)}$ tends to converge to a large value in the region $t_1/t_2=\pm 2$, indicating increased quantum fluctuations. }
  \label{fig:entropy_distribution}    
\end{figure*}

\subsection{Hard-Core Bosons, $U\rightarrow \infty$ limit:  mapping to $J_1$-$J_2$-model }
\label{hardcore_bosons}
In the limit $U\rightarrow \infty$, the occupation of bosons per site is limited to $\max_i\{n_i\}=1.$ Hence, we can map the Bose--Hubbard model to a spin-1/2 system.  One way to do that is following the   Holstein–Primakoff transformation \cite{Holstein}
\begin{equation}
\hat{S}^ {+}=\sqrt{2s}\sqrt{1-\frac{\hat{n}}{2s}}\hat{b}\approx \hat{b},
\label{eqn:HP}
\end{equation}
where $s$ is the spin of the particles,   and similar for $\hat{S}^ {-}$.  In the limit $U \rightarrow \infty$,  the relation in Eq. {\!}(\ref{eqn:HP})  becomes exact. With this transformation, the BH model with NNN-tunneling maps to 
\begin{eqnarray}
\hat{H}
&\approx&-t_1\sum_{\langle i,j \rangle}\Big(\hat{S}^{-}_i\hat{S}^+_j+h.c.\Big)-t_2\sum_{\langle \langle i,j \rangle \rangle}\Big(\hat{S}^-_i\hat{S}^+_j+h.c.\Big),\nonumber\\
&=&J_1\sum_{\langle i,j \rangle}\left(\hat{S}^x_i\hat{S}^x_j+\hat{S}^y_i\hat{S}^{y}_j\right)+J_2\sum_{\langle \langle i,j \rangle \rangle}\left(\hat{S}^x_i\hat{S}^x_j+\hat{S}^y_i\hat{S}^y_j\right),\nonumber\\
\end{eqnarray}
with $\hat{S}^{x}_i=\frac{\hat{S}^++\hat{S}^-}{2}$, $\hat{S}^{y}_i=\frac{\hat{S}^+-\hat{S}^-}{2i}$,  $J_1=-2t_1$, and $J_2=-2t_2$. The transformed Hamiltonian is the well-known $XX$-model with NN- and NNN-interactions also commonly known as the $J_1$-$J_2$ model. From  earlier works \cite{sato-et-al1,sato-et-al2},  it is known that for  $0\leq |J_1|/J_2 \leq 4$, there exist three regions, namely, Tomonaga-Luttinger liquid (TLL) phase, even- or odd-dimer phase, and vector chiral phase.

\begin{figure*}
  \includegraphics[width=17cm]{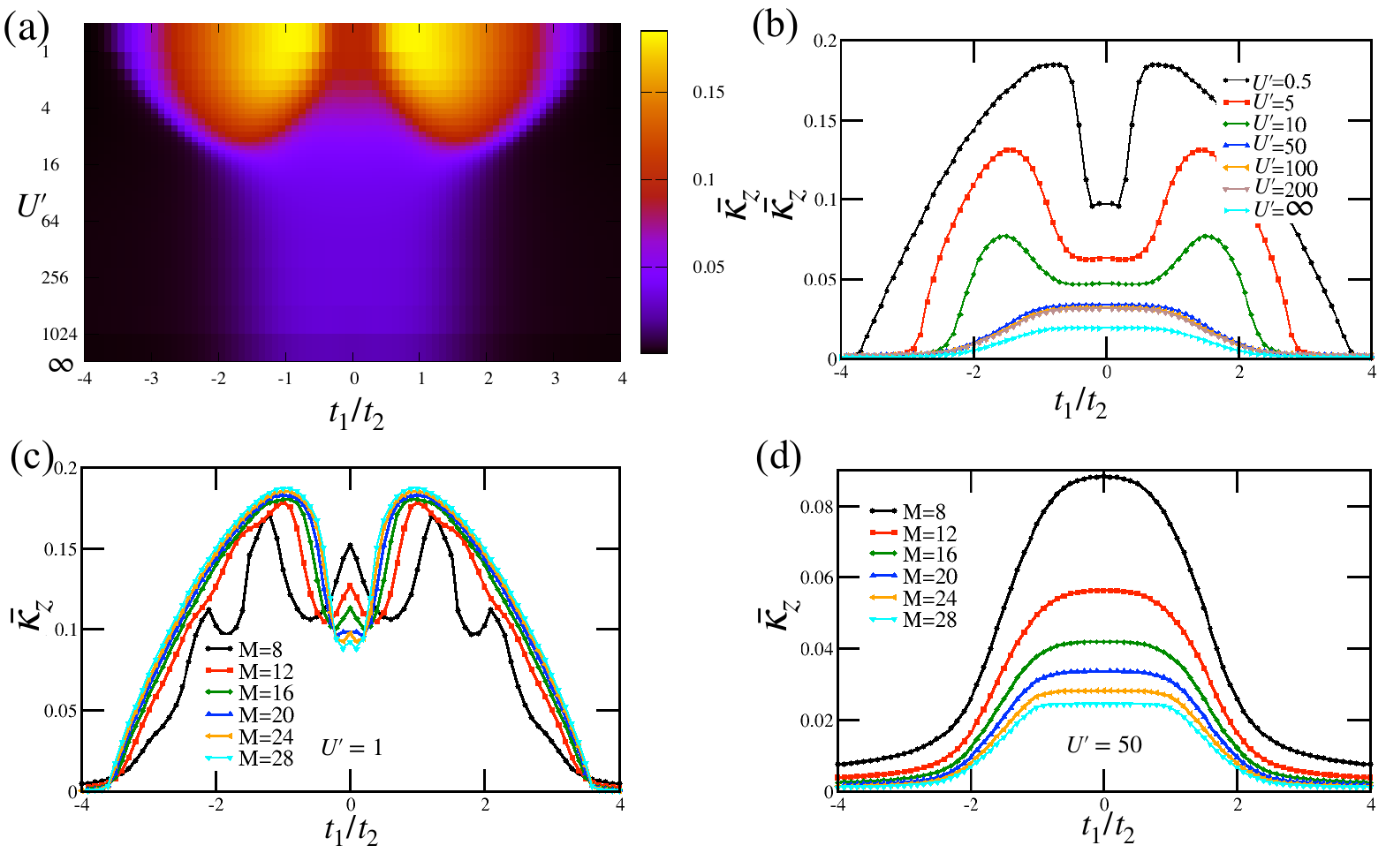}
  \caption{(a) Behavior of chiral order parameter $\bar{\kappa}_{z}$  for  system size $N=10$, $M=20$. (b) Scaling of $\bar{\kappa}_{z}$ with  different values of onsite interaction  $U$, and for the same system size as in panel (a). In panel (c) and (d), we provide a finite-size scaling of $\bar{\kappa}_{z}$ with system size $M= 8, 12, 16, 20, 24$, and 28, for two values of onsite interactions, $U=1$ and 50, respectively. The finite-size scaling analysis shows with the increasing system size $\bar{\kappa}_{z}$ tends to converge in all the considered regions. At small $U$, it remains finite in the region $|t_1|/t_2\lesssim 4$, corresponding to the spiral phase at $U\rightarrow0$ whereas for large $U$ this region shrinks to $|t_1|/t_2 \lesssim 2$.}
  \label{fig:chiral_order_parameter}    
\end{figure*}

\section{Results}
\label{sec:results} 
In our work, our main focus will be on the intermediate values of $U$. In other words,  we wish to find how different orders in the system change in presence of finite but nonzero $U$. Towards that aim, we compute a list of quantities and order parameters, using numerically exact diagonalization as  well as tensor-network methods by employing  the density matrix normalizing group \cite{itensor} technique.  Here,  for all computational purposes we consider the system  at half-filling, $N=M/2$,  and we set $t_2=-1$.   Additionally, we denote  $U'=U/|t_2|$ .

\subsection{Population in   momentum modes}
\label{sec:population}
In Fig. {\!}\ref{fig:density}(a), we plot the  maximum  bosonic  density $\eta=\max_q n_q/N$ (with $n_q=1/M\sum_{ij}e^{-iq(i-j)}\langle {\hat{b}_i}^\dagger \hat{b}_j \rangle$)  in the GS.  
In panel (b), we plot the corresponding momentum mode with maximum population, $q_{\max}$. $\eta$ is  also known as the structure factor of the system, and a nonzero value of it in the thermodynamic limit  guarantees presence of long-range order (LRO) at the wave vector $q_{\max}$. As here we are considering OBC, $q$ is no longer a good quantum number and it can take any continuous value in between  $q\in (-\pi, \pi]$, which helps to alleviate incommensurability effects appearing in the frustrated system due to finite system size. In our analysis, we took $N_q=10^3$  values of $q$ (with a difference $\Delta_q\pi$). 

  From Fig. {\!}\ref{fig:density}(a), 
we can see that the GS population density is maximum near $U\rightarrow 0$ and $|t_1|/t_2\sim 4$. With the increase of $U$, bosons no longer condense into a single mode  and  $\eta$ becomes minimal approximately in the region $1<|t_1|/t_2<2$.  
Moreover,  the maximally populated mode $q_{\max}$   broadly divides the region $-4\leq t_1/t_2<0$ (equivalently, $0<t_1/t_2\leq 4$) into two parts:   a region with predominant (i) commensurate (C) order $q_{\max}=0$ ($q_{\mathrm{max}}=\pi$) and (ii) an incommensurate (IC) order  $0 < |q_{\mathrm{max}}|<\frac{\pi}{2}$ ($\frac{\pi}{2} < |q_{\mathrm{max}}|<\pi$). We denote the transition  by a dashed yellow (blue) line in Fig. {\!}\ref{fig:density}(b), which is also known as the Lifshitz line \cite{sato-et-al1,sato-et-al2}. Comparing with the behavior of $\eta$, $q_{\max}$ indicates the pitch vector of dominating phase order, but not necessarily (quasi-)LRO.

\begin{figure*}
  \includegraphics[width=17cm]{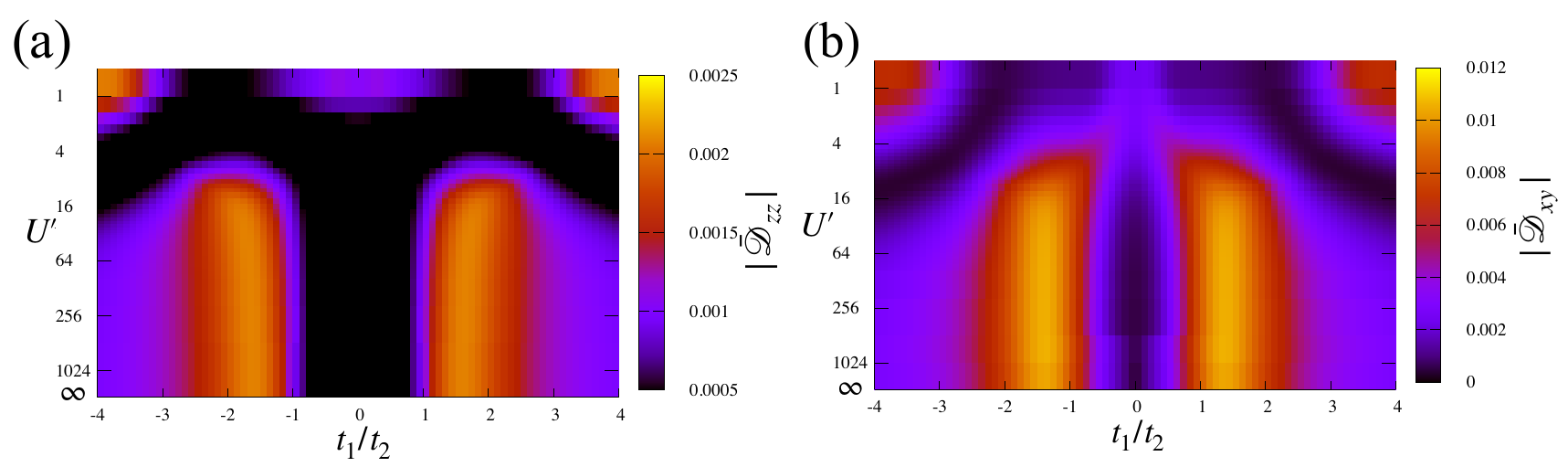}
  \caption{Dimer correlators  obtained for the GS of the model.   (a) Dimer $zz$ correlator $|\bar{\mathcal{D}}_{zz}|$ and  (b) dimer $xy$ correlator $|\bar{\mathcal{D}}_{xy}|$ for $N=10$, $M=20$.  For a large portion in the parameter space, both the quantities remain significantly low. However,   in the region $1<|t_1|/t_2<3$, i.e., in the region with strong quantum fluctuations,  both of them possess high values.} 
  \label{fig:dimer_order_parameter}    
\end{figure*}

Apart from the maximum population of the modes, the distribution of the particles among all the modes can give us a more detailed structure of the GS of the Hamiltonian. For that purpose, we compute the  quantity $\mathcal{S}_q=-\sum_q \rho_q \ln \rho_q$, where $\rho_q=n_q \Delta_q$. Here, the factor  $\Delta_q$ ensures normalization $\sum_q \rho_q=1$. 
 We plot the behavior of $\mathcal{S}_q/\max(\mathcal{S}_q)$  in Fig. {\!}
 \ref{fig:entropy_distribution}, 
 where $\max(\mathcal{S}_q$) is the theoretical maximum obtained at  $\rho_q=1/N_q$. $\mathcal{S}_q$ reflects   the structures  predicted by $\eta$.
 
  As discussed earlier, with  increase of the contact interaction $U$, the GS starts populating other momentum modes, resulting in a high value of  $\mathcal{S}_q$ for the region $1<|t_1|/t_2<2$.  A high value of  $\mathcal{S}_q$  in this context implies stronger quantum fluctuations and reduction of phase order, which is consistent with the behavior observed  in the hard-core boson limit, where $\mathcal{S}_q$ remains significantly high for the dimer phase and  becomes maximum   at   $|t_1|/t_2\approx 2$; see the scaling  of $\mathcal{S}_q/\max(\mathcal{S}_q)$ with $U$ presented in Fig. {\!}\ref{fig:entropy_distribution}(b). Additionally, we provide a finite-size scaling of $\mathcal{S}_q/\max(\mathcal{S}_q)$ with $M$ in Fig. {\!} \ref{fig:entropy_distribution}(c) that suggests $\mathcal{S}_q/\max(\mathcal{S}_q)$ tends to converge in the region $t_1/t_2=\pm 2$ even at moderately high system size. As we will see below, it is in these regions where the GS of the system  comprises strong quantum fluctuations, leading to quantum phases without classical analog,  in particular dimerized phases.


\subsection{Vector chiral order}
\label{chiral_order}
To further understand the nature of the quantum phases appearing in the model, we consider as order parameter the chiral correlator, which  measures chiral correlations between sites that are separated by a distance $\Delta$ \cite{Hikihara2001, Schmied}, given by 
\begin{equation}
\kappa_{z}^{\Delta}=\sum_{j=1}^{M-1-|\Delta|}\langle \kappa^{z}_{j}\kappa^{z}_{j+\Delta}\rangle. 
\end{equation}
Here   $\kappa_j^{z}=\frac{1}{2i}({\hat{b}_j}^\dagger\hat{b}_{j+1}-{\hat{b}_{j+1}}^\dagger\hat{b}_{j})$, which using the spin-1/2 operators can also be written as $\kappa_j^{z}= \left(\mathbf{S}_{j}\times \mathbf{S}_{j+1}\right)^{z}$. 
In addition, we also define the average of $\kappa_{z}^{\Delta}$ over $\Delta$,

\begin{equation}
  \bar{\kappa}_{z} =
       \frac{1}{2M-3}\sum_{\Delta=-(M-2)}^{M-2}\frac{\kappa_{z}^{\Delta}}{M-1-|\Delta|}        .
\end{equation}

A nonzero value of  $\bar{\kappa}_{z}$  in the thermodynamic limit certifies the presence of long-range chiral order in the system. We present the behavior of $\bar{\kappa}_{z}$ in Fig.  {\!} {\ref{fig:chiral_order_parameter}}(a), for  $M=20$. Near $U\rightarrow0$,  $\bar{\kappa}_z$ remains significantly high which is reminiscent of the classical spin spiral phase appearing in this region.
   As we increase  $U$ further, we can see $\bar{\kappa}_{z}$ can distinguish the predominating commensurate phase order  ($q_{\max}=0$ and $\pi$) at $|t_1|/t_2\gtrsim 2$   as it  remains significantly low.

 However, the relatively larger value  for the regime $0<|t_1|/t_2<2$ needs more careful interpretation. For instance,  in  the hard-core boson limit,  even in some regions of  the  dimer phase ($1< |t_1|/t_2 < 2$)$,  \bar{\kappa}_{z}$ takes a low but finite value.  In this case, a relatively higher growth is observed   around  $|t_1|/t_2\approx 1$, and $\bar{\kappa}_{z}$ starts saturating, which indicates the vector chiral ordered phase in the model \cite{Schmied}. Figure \ref{fig:chiral_order_parameter}(b) suggests   a similar behavior  is    observed  even when the system is away from the hard-core boson limit, where    $\bar{\kappa}_{z}$ grows fast and eventually  saturates to  a high value. We provide a finite-size scaling of $\bar{\kappa}_{z}$ with the system size $M$ in Figs. {\!}\ref{fig:chiral_order_parameter}(c) and (d) for two values of the onsite interaction, $U=1$ and $50$, respectively. We note that with increasing the system size $M$, $\bar{\kappa}_{z}$ tends to converge and the rate of convergence is higher for low values of $U$. Moreover,  for small $U$, even at higher $M$ $\bar{\kappa}_{z}$  remains nonzero for almost all considered values of $|t_1|/t_2$. In contrast, for large onsite interaction, $\bar{\kappa}_{z}$  decreases with increasing system size $M$ and remains significant only for the region $|t_1|/t_2<2$. 
 
\begin{figure*}
  \includegraphics[width=16cm]{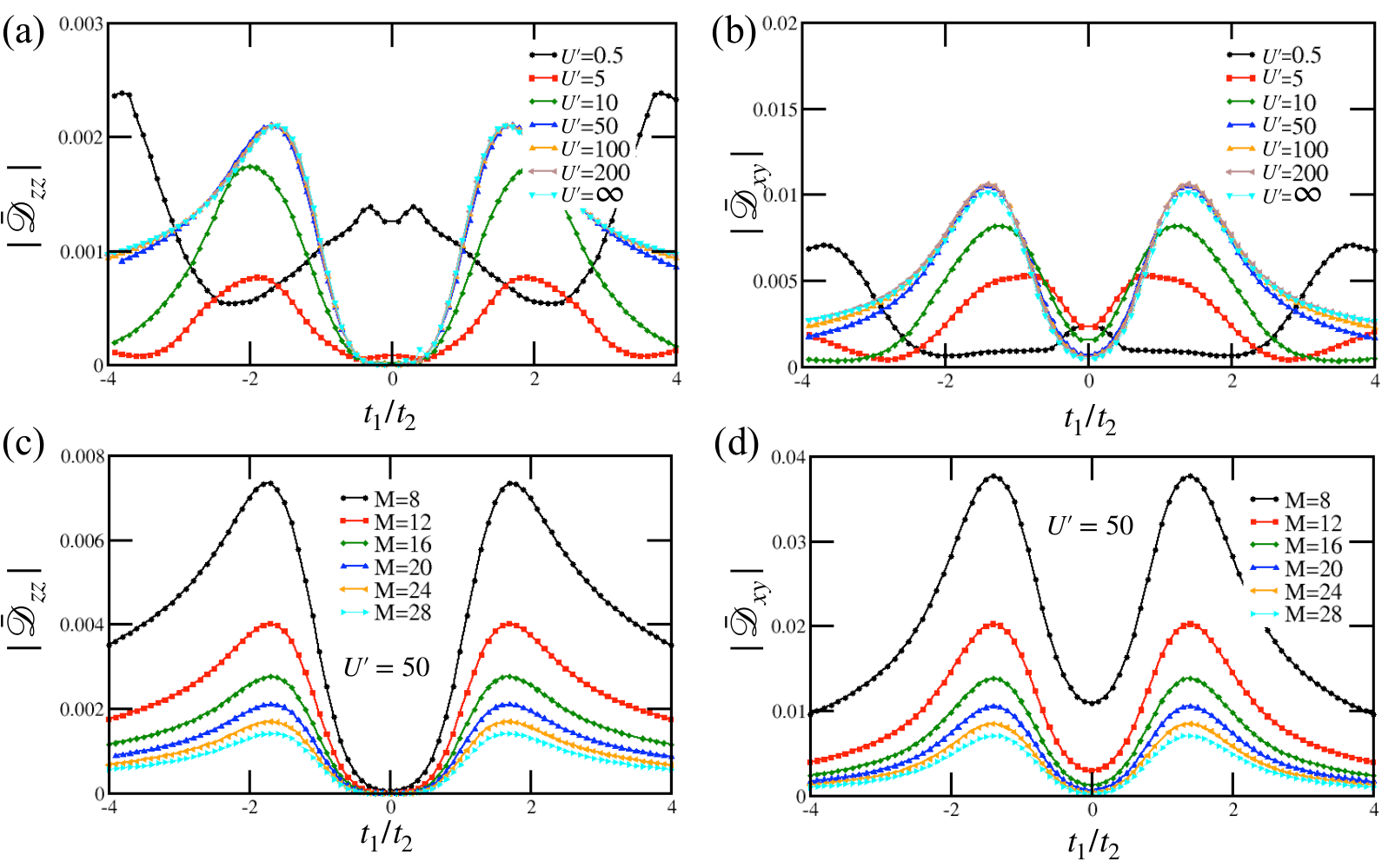}
  \caption{Scaling of  dimer correlators with onsite interaction $U$ and system size $M$. Change of (a) $\bar{\mathcal{D}}_{zz}$ and (b) $\bar{\mathcal{D}}_{xy}$ with onsite interaction $U$, for fixed  $N=10$, $M=20$.  (c) and  (d), finite-size scaling of both  quantities with the system size $M$ for $M=8, 12, 16, 20, 24, 28$, for fixed $U'=50$. The scaling suggests for large $M$, both remain significantly low except for the region with strong quantum fluctuations, i.e., $1<|t_1|/t_2<3$.}
  \label{fig:dimer_order_parameter_scaling}    
\end{figure*}

\subsection{Dimer order}
\label{dimer_order}
Further information on the GS behavior can be attained from  the dimer order parameters, which can be defined as follows \cite{ sato-et-al1,sato-et-al2},   
{\small
\begin{equation}
\begin{split}
\mathcal{D}^{xy}_{j}
&= \frac{1}{2}\langle \left({\hat{b}_j}^\dagger\hat{b}_{j-1}+{\hat{b}_{j-1}}^\dagger\hat{b}_{j}\right)-\left({\hat{b}_j}^\dagger\hat{b}_{j+1}+{\hat{b}_{j+1}}^\dagger\hat{b}_{j}\right)\rangle, \\
\end{split}
\end{equation}
}
and 
\begin{equation}
\begin{split}
\mathcal{D}^{zz}_{j}&=\langle\left(\frac{1}{2}-\hat{n}_j\right)\left(\frac{1}{2}-\hat{n}_{j-1}\right)-\left(\frac{1}{2}-\hat{n}_{j+1}\right)\left(\frac{1}{2}-\hat{n}_{j}\right)\rangle.
\end{split}
\end{equation}
Similar to the chiral order parameter, we define the dimer correlators  as
 
\begin{equation}
\mathcal{D}_{xy}^{\Delta}=\sum_{j=2}^{M-1-|\Delta|}\langle \mathcal{D}^{xy}_{j}\mathcal{D}^{xy}_{j+\Delta}\rangle,\quad \mathcal{D}_{zz}^{\Delta}=\sum_{j=2}^{M-1-|\Delta|}\langle \mathcal{D}^{zz}_{j}\mathcal{D}^{z}_{j+\Delta}\rangle,
\end{equation}
 
and 
\begin{eqnarray}
\bar{\mathcal{D}}_{xy}&=&\frac{1}{2M-5}\sum_{\Delta=-(M-3)}^{M-3}\frac{\mathcal{D}_{xy}^{\Delta}}{M-2-|\Delta|},\nonumber\\
\quad \bar{\mathcal{D}}_{zz}&=&\frac{1}{2M-5}\sum_{\Delta=-(M-3)}^{M-3}\frac{\mathcal{D}_{zz}^{\Delta}}{M-2-|\Delta|}.
\label{eqn:dimer_order_paramter}
\end{eqnarray}

We present the behavior of  ${\bar{\mathcal{D}}}_{zz}$ and ${\bar{\mathcal{D}}}_{xy}$ for different values of  $U$ and $t_1/t_2$ in Figs.  {\!}\ref{fig:dimer_order_parameter}(a) and (b), respectively. For low values of $U$, both  quantities remain low for the maximum regions in the considered parameter space, in agreement with an absence of dimer order in that regime.  However, relatively high values of both the measures can be  observed for a small region around $q_{\max}\rightarrow 0$ and $\pi$.

 On the contrary, in the hard-core boson limit,  they can be considered as a good identifier of the quantum phase boundaries: in the dimer phase, both quantities attain maximum values, while they  remain significantly low for the vector chiral  and TLL phases.  
 At   high but finite  values of $U$, the  characteristics of both ${\bar{\mathcal{D}}}_{zz}$ and ${\bar{\mathcal{D}}}_{xy}$ remain  similar to that observed for the hard-core boson limit and  the GS again  possesses  finite and high values of dimer correlators for the region   $1<|t_1|/t_2<3$ [see Figs. {\!} \ref{fig:dimer_order_parameter_scaling}(a) and (b)]. Moreover, similar to vector chiral phase, the quantities remain significantly low in the region $0<|t_1|/t_2<1$. The behavior  of  ${\bar{\mathcal{D}}}_{xy}$ and  ${\bar{\mathcal{D}}}_{zz}$ in the region $3<|t_1|/t_2<4$ remain intermediate to the previous two regimes.  A  finite-size scaling  for $U=50$ [see Figs. {\!]
\ref{fig:dimer_order_parameter_scaling}(c) and (d)] suggests at large $M$   both the quantities tend to converge in all regions:  for $|t_1|/t_2<1$ and $|t_1|/t_2>3$, they become significantly low. In contrast, for  $1<|t_1|/t_2<3$ both of them possess higher  values, indicating strong quantum fluctuations leading to dimerization. We provide analytical forms of the scaling of both the quantities with the  system size $M$ for  the exact dimerization point, $|t_1|/t_2=2$ [see Appendix {\!}\ref{AppendixB}],  given by 
\begin{eqnarray}
|\bar{\mathcal{D}}_{zz}|&=&\frac{1}{(2M-5)}\frac{M-2}{16(M-3)}, \nonumber\\
|\bar{\mathcal{D}}_{xy}|&=&\frac{1}{8(M-3)}.
\end{eqnarray} 
For example, for $M=20$, we get $|\bar{\mathcal{D}}_{zz}|=0.00189$ and $|\bar{\mathcal{D}}_{xy}|=0.00735$, see  Fig.       	{\!}\ref{fig:dimer_order_parameter_scaling}(a) and (b).
We summarize the behavior of the order parameters obtained above in the schematic Figs.{\!} \ref{fig:schematic1}(b). In the forthcoming section, we make a comparative study of these findings with the behavior of the entanglement properties in the GS of the system.

\begin{figure*}
  \includegraphics[width=16cm]{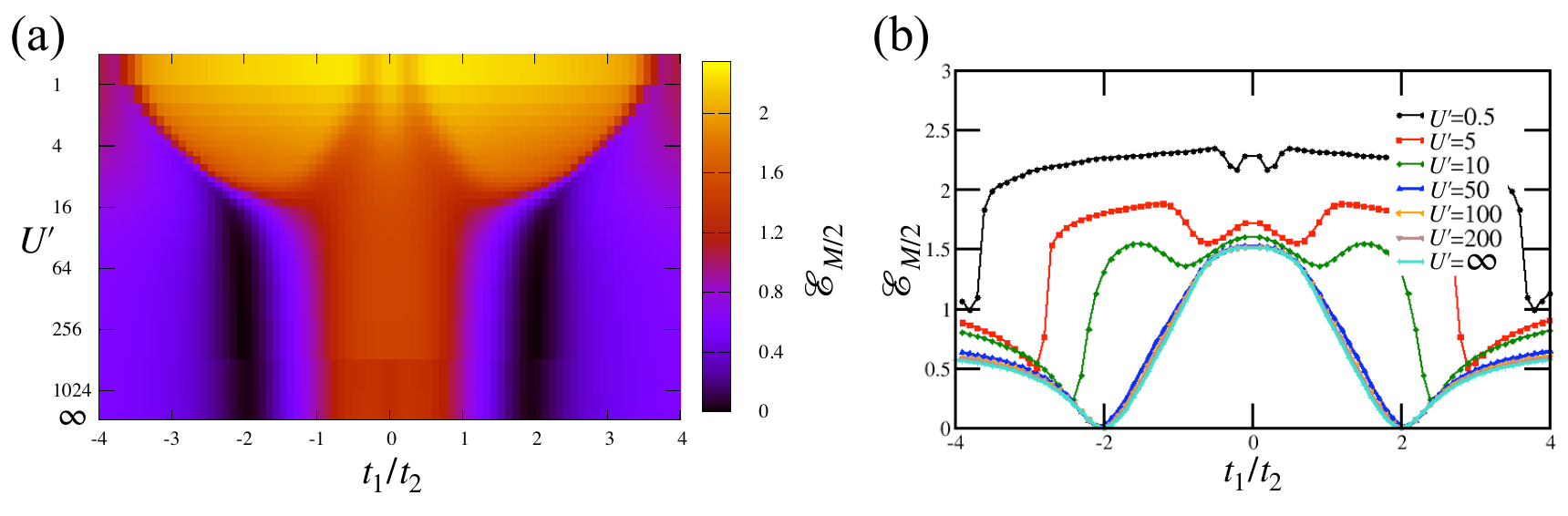}
  \caption{(a)  Behavior of half-chain entanglement entropy ($\mathcal{E}_{M/2}$) for  $N=10$, $M=20$.  For the same system size, we plot cuts at fixed   onsite interaction $U$ in panel (b). We note that for most of the regions in the considered parameter space, the behavior of $\mathcal{E}_{M/2}$ remains compatible with $\bar{\kappa}_z$.  In contrast, at large $U$, in the region with strong quantum fluctuations, i.e., $1<|t_1|/t_2<3$, where the dimer orders dominate, $\mathcal{E}_{M/2}$ remains significantly low. In particular, in the hard-core boson limit at $|t_1|/t_2=2$ GS becomes a product of dimers and  $\mathcal{E}_{M/2}$ is exactly zero.} 
  \label{fig:bipartite_entanglement_scaling_U}    
\end{figure*}

\section{Entanglement properties}
\label{sec:entanglement}
In this second part of our work,  we analyze the behavior of both bipartite and multipartite entanglement  obtained for the GS of the model. 

\subsection{Bipartite entanglement}
We start our discussion with the analysis of bipartite entanglement present in the GS of the system. As a measure of bipartite entanglement, we consider the half-chain  entanglement entropy,  defined as 
\begin{eqnarray}
\mathcal{E}_{M/2}=-\Tr(\rho \ln \rho),
\end{eqnarray}
where $\rho=\Tr_{M/2+1\dots M}({_M}|\Psi\rangle\langle \Psi|_M)$ is the reduced density matrix consisting of  half of the chain, and plot its behavior in  Fig. {\!}\ref{fig:bipartite_entanglement_scaling_U}(a).

 For a large portion of the considered parameter regime, the behavior of bipartite entanglement $\mathcal{E}_{M/2}$ remains compatible with that of the chiral correlator, $\bar{\kappa}_{z}$, as shown in Fig. {\!} \ref{fig:chiral_order_parameter}(a). Similar to   $\bar{\kappa}_{z}$, $\mathcal{E}_{M/2}$ remains significantly high for low values of contact interaction $U$ and for almost all values of $t_1/t_2$ in the considered parameter region. With the increase of $U$, $\mathcal{E}_{M/2}$ shows interesting behavior. For instance, in the hard-core boson limit, it can distinguish three phases clearly. In the chiral phase ($|t_1|/t_2<1$), similar to $\bar{\kappa}_{zz}$, $\mathcal{E}_{M/2}$ increases monotonically and attains its maximum value. In contrast, for a wide region  of  the dimer phase ($1<|t_1|/t_2<3$) $\mathcal{E}_{M/2}$ remains considerably low and at the exact dimerization point,  $|t_1|/t_2=2$, $\mathcal{E}_{M/2}$ attains its minimum value $\mathcal{E}_{M/2}=0$. In the region $|t_1|/t_2>3$,  $\mathcal{E}_{M/2}$ again shows a monotonic growth and saturates to a lower value than that of the chiral phase. Away from the hard-core boson limit,  for a finite but large $U$, the behavior remains qualitatively similar. This suggests in terms of bipartite entanglement, the phase boundaries obtained for hard-core boson limit translate up to a large but finite $U$.

\begin{figure*}
  \includegraphics[width=16cm]{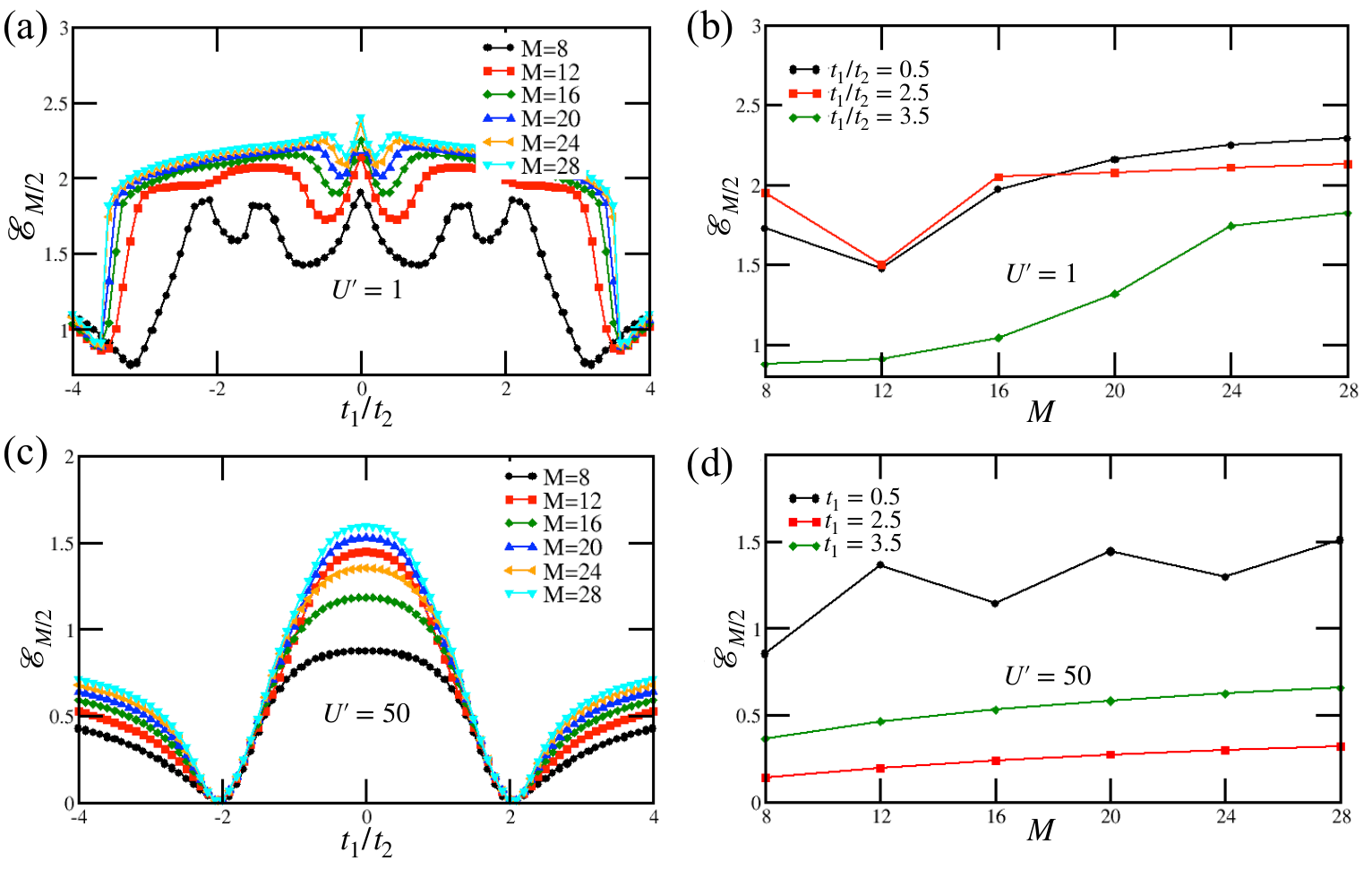}
  \caption{Scaling of $\mathcal{E}_{M/2}$ with onsite interaction ($U$) and system size ($M$). (a)  and (c) depict the scaling of $\mathcal{E}_{M/2}$ with $M$ for two values of onsite interactions, $U=1$ and $U=50$, respectively. (b) and (d) show a more detailed picture of the rate of convergence of  $\mathcal{E}_{M/2}$ observed in those plots for different values of frustration $|t_1|/t_2$.  At low $U$, bipartite entanglement seems to converge to  nonzero values for almost all considered values of $|t_1|/t_2$. The same is true when we consider large $U$, except for the regions around  $|t_1|/t_2=2$.}
  \label{fig:bipartite_entanglement_scaling_N}    
\end{figure*}

 \begin{figure*}
  \includegraphics[width=17cm]{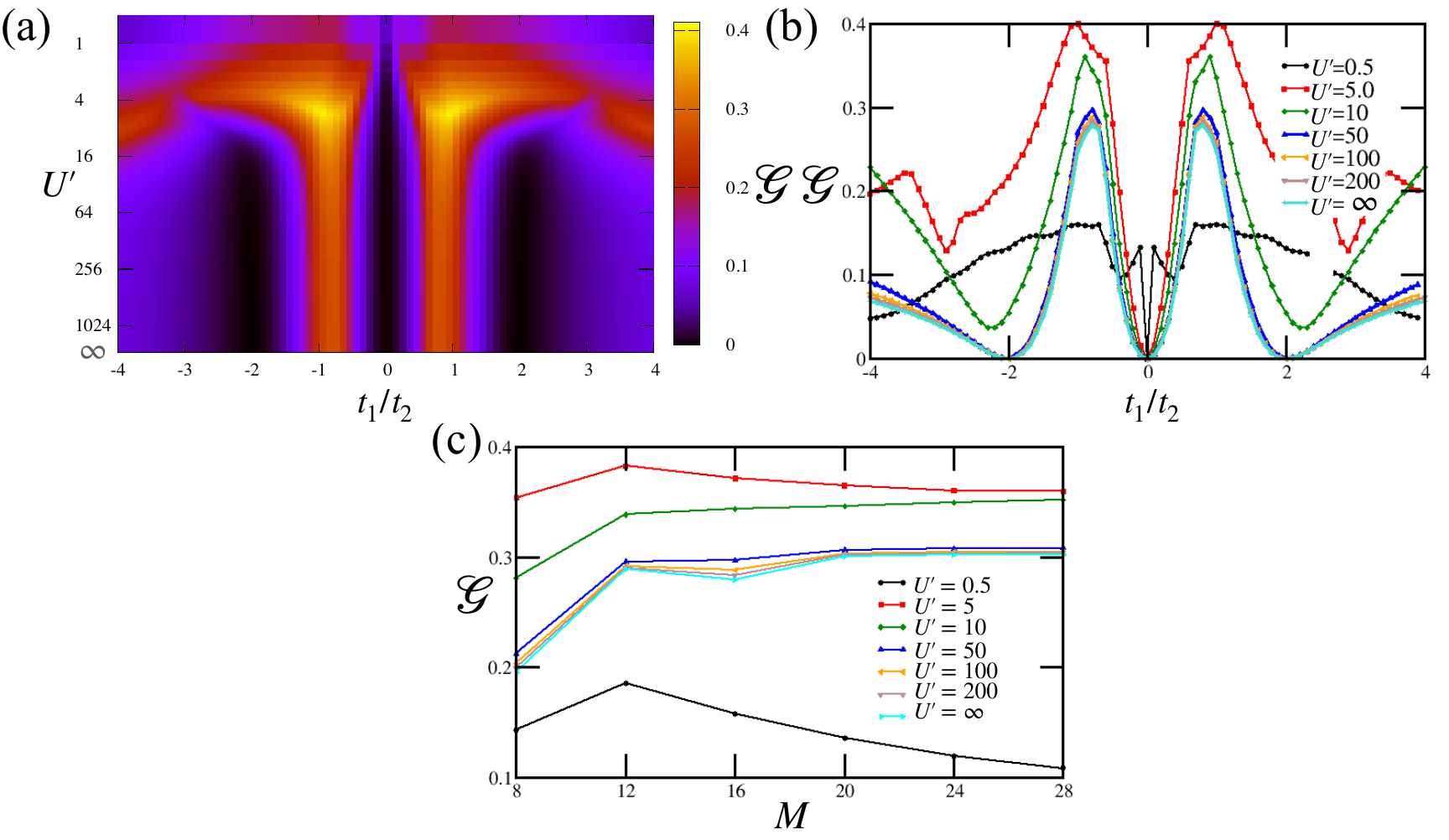}
  \caption{(a) Behavior of generalized geometric measure ($\mathcal{G}$) with onsite interaction and NN and NNN hoppings. Data for $N=8$, $M=16$. In panel (b), we plot the cuts  of $\mathcal{G}$ against $U$ for the same system size. For low $U$, except for the point $t_1/t_2=0$, $\mathcal{G}$ remains nonzero for the entire considered region. In contrast, for large $U$,  except for the region adjacent to $|t_1|/t_2\approx 1$, $\mathcal{G}$ remains low for almost all values of the considered parameter region. In particular,  in the hard-core boson limit $\mathcal{G}$ becomes zero also at the exact dimerization point $|t_1|/t_2=2$. Interestingly, $\mathcal{G}$ attains its peak near $|t_1|/t_2\approx 0.8$ which  roughly marks the onset point of the vector chiral phase that appears in the hard-core boson limit of the model. Panel (c) shows the scaling of $\mathcal{G}$ with system size $M$  (with $M=8, 12, 16, 20, 24$, and 28), for different $U$ values at $t_1/t_2=0.8$. One can see that   $\mathcal{G}$ tends to converge to a finite value with the increase of system size, except for small $U$. }
  \label{fig:GGM_entanglement}    
\end{figure*}

A finite-size scaling analysis in Fig. {\!}\ref{fig:bipartite_entanglement_scaling_N} depicts how $\mathcal{E}_{M/2}$  scales with system size $M$ for small [$U'=1$, panel (a)  and (b)] and large  [$U'=50$, panel (c) and (d)] values of $U$. In all the cases, $\mathcal{E}_{M/2}$ tends to converge fast with the increase of system size $M$.  For small $U$,  $\mathcal{E}_{M/2}$  remains nonzero for almost all  values of $t_1/t_2$. At large $U$, the same behavior is observed  except for the  point, $|t_1|/t_2=2$, and some regions adjacent to that.

\subsection{Genuine multipartite entanglement}
As mentioned previously, in general the computation of multipartite entanglement even for a pure quantum state is a difficult task, and   in the literature 
there exist several inequivalent definitions and measures of it; see, e.g.,  \cite{ggm1,ggm2,ggm3,ggm4,chiara2}. In our case, we mainly focus on genuine multipartite entanglement (GME) of the system, which is defined as follows. 
 ``An $M$-party pure quantum state, $|\Psi\rangle_M$, is said to be genuinely multipartite entangled  if it cannot be written as a product in any possible bipartition of the state".  As a measure of GME, we consider the generalized geometric measure (GGM), which is a computable measure and quantifies the distance of the given $M$-party state from the set of states that are not genuinely multipartite entangled. Mathematically, this can be expressed as
 \begin{equation}
 \mathcal{G}(|\Psi\rangle_M)=1-\max_{|\zeta\rangle \in \mathcal{S}_{\zeta}}|\langle \zeta|\Psi\rangle_M|^2,
 \end{equation}
 where $\mathcal{S}_{\zeta}$ is the set of states consisting of nongenuinely multipartite entangled states. One can show that an equivalent expression of the above equation is 
  \begin{equation}
 \mathcal{G}(|\Psi\rangle_M)=1-\max_{A:B}\lambda_{A:B}^2,
 \end{equation}
 where $\lambda_{A:B}$ is the maximum Schmidt coefficient across the bipartition $A:B$  with $A, B, \in \{1,2,\dots, M \}, A \cap B= \emptyset, A \cup B=M$, 
 and the maximization has been performed over all possible bipartitions of the state. Because we can decompose  any given pure state $|\Psi\rangle_M$ in Schmidt form and in principle get $\lambda_{A:B}$ from that, the employed measure does not depend on the dimensionality or underlying lattice structure and can thus be used for a wide range of models, including in higher dimensions \cite{hsd,ssr0,ggm6} or with longer-ranged interactions \cite{ssr1}.

We are now ready with the necessary tools to analyze the behavior of GME in the system. Figure  \ref{fig:GGM_entanglement}(a) depicts the behavior of $\mathcal{G}$ obtained for the GS of the model in the $U'-t_1/t_2$ plane.  From the plot, we note that for low values of onsite interaction and $|t_1|/t_2>1$ the behavior of the GME remains characteristically similar to that of the bipartite entanglement and chiral correlator discussed above, and $\mathcal{G}$ attains a high value. In that region, the optimum value of Schmidt coefficient $\lambda_{A: B}$ is obtained for contiguous blocks.

 In contrast,  for $|t_1|/t_2<1$, i.e., when NNN hopping dominates, unlike the previous two quantities,  $\mathcal{G}$ reduces to very low value and the  bipartition with $A \in 2, 4,6, \dots,  N$, i.e., all even-sites (or equivalently $B \in 1, 3, 5, \dots, N-1$, i.e., all odd-sites) yields maximum $\lambda_{A:B}$.  As the value of onsite interaction is increased further, except for a small region around $|t_1|/t_2\sim 1$, the global entanglement shared between the particles remains small in all the regions of the considered parameter space.   Similar to bipartite entanglement, for a large fraction of the  region with dominating dimer order  ($1<|t_1|/t_2<3$),  $\mathcal{G}$  remains very low.  This suggests  in the region where the quantum particles remain maximally spread or delocalized in all the momentum modes, and thus result in a high value of  $\mathcal{S}_q$,  quantum entanglement remains very low or localized. At the exact dimerization point (for $U\rightarrow \infty$), $|t_1|/t_2=2$, where  $\mathcal{S}_q$ is maximum, similar to BE,  GME also vanishes, as the state separates into a product of dimers.  Interestingly, from  figure \ref{fig:GGM_entanglement}(b) we can see that the peak of $\mathcal{G}$ at $|t_1|/t_2\approx 0.8$ appears close  to the onset point of the vector chiral phase of the model appearing in the hard-core boson limit. However, unlike BE, GME remains   low  for almost all the regions with dominating chiral correlation ($|t_1|/t_2<1$), which is very different from the behavior of bipartite entanglement observed in that region. =In this region, the system factorizes as $|\Psi\rangle_M)\approx |\phi\rangle_{\text{even sites}} \otimes |\tilde{\phi}\rangle_{\text{odd sites}}$, and
thus  multipartite entanglement becomes zero. A finite-size scaling analysis in Fig. {\!}\ref{fig:GGM_entanglement}(c) shows $\mathcal{G}$ converges with $M$ fast even for a moderately high value of $U$.  We summarize the behavior of entanglement properties of the model in  the  schematic presented in Fig. {\!}\ref{fig:schematic1}(b).

Therefore,  though in some portions of the considered region the behavior of quantum entanglement remains compatible with different order parameters computed for the GS of the system, for a large part of the parameter space, a significant difference  can also be observed.  In addition to this,  depending on the strength of frustration and onsite interaction, even different forms of quantum entanglement (bipartite and multipartite) exhibit distinct behavior. As this shows, the characterization of quantum entanglement in the model demands separate attention from the analysis of ordering properties.

\section{Conclusion}
\label{conclusion}

In this work, we considered the frustrated Bose--Hubbard model with  NN and NNN hoppings and examined the quantum properties of the GS of the model. Starting with two limit cases, free bosons and hard-core bosons, we analyzed how different order parameters change with the interplay of frustration and finite onsite interaction. We observed that for the chiral and dimer correlations, the behavior observed in the hard-core boson limit  approximately  translated up to a moderately large value of $U$. Similar behavior was also reported for the population of quantum particles in different momentum modes. We then analyzed the bipartite and multipartite entanglement properties of the system and compared those with the conventional order parameters mentioned above. We found  half-chain entanglement entropy  remains high for regions with a high value of chiral correlation. However, for  the generalized geometric measure, this remains true only for the region with dominating NN hopping.  Along with this, almost all regions with dominating dimer order in the system yield a very low value of both bipartite and genuine multipartite entanglement.

 Hence, our analysis reveals that  entanglement properties of the system often do not manifest in the behavior of the conventional order parameters computed for the GS of the system and thus deserve separate attention. Moreover, strong quantum fluctuations do not necessarily  imply multipartite entanglement. In the present model, e.g., the system resolves strong frustration by tightly  binding neighboring sites into dimers and thus due to the monogamy of entanglement, suppressing the distribution of quantum entanglement among a large number of parties. 

In the future, it will be interesting to perform similar analyses on other models known for complex quantum mechanical phase diagrams, e.g., those with long-range density-density interactions \cite{ref_BH2,conclusion2} or frustrated models in two dimensions \cite{Eckardt2,conclusion3,conclusion4}, as well as to compare results to other entanglement measures. 

\acknowledgements
We acknowledge support by the ERC Starting Grant StrEnQTh (project ID 804305), Provincia Autonoma di Trento, and by Q@TN, the joint lab between University of Trento, FBK-Fondazione Bruno Kessler, INFN-National Institute for Nuclear Physics and CNR-National Research Council. We thank Soumik Bandyopadhyay for  reading the manuscript and providing useful suggestions. We also acknowledge the use of  iTensor C++ library for the DMRG computations performed in this work \cite{itensor}.

\appendix
\begin{widetext}

\section{Analytical values of dimer correlator for perfect Dimerized state}
\label{AppendixB}
The perfect triplet dimer for the hard-core limit assumes the form 
\begin{equation}
|D\rangle=\Pi_{i=1}^{M/2}\left(\frac{\hat{b}^\dagger_{2k-1}+\hat{b}^\dagger_{2k}}{\sqrt{2}}\right)|0\rangle,
\label{eq:perfect_dimer} 
\end{equation}
where we choose the bonds between the dimers to be between odd-even sites.

We write the single-site dimer order parameter as 
\begin{equation}
\hat{\mathcal{D}}^{xy}_{k}=\left(\hat{j}_{k}-\hat{j}_{k+1}\right),
\end{equation}
where $\hat{j}_{k}=\frac{1}{2i}(\hat{b}^\dagger_{k}\hat{b}_{k-1}+\hat{b}^\dagger_{k-1}\hat{b}_{k})$ denotes the current between site $k$ and site $k-1.$
With the above dimer form we have the following identities:
\begin{eqnarray}
\langle D|\hat{\mathcal{D}}^{xy}_k\hat{\mathcal{D}}^{xy}_{k}|D\rangle &=&-\frac{3}{8},\\
\langle D|\hat{\mathcal{D}}^{xy}_{k}\hat{\mathcal{D}}^{xy}_{k+1}|D\rangle&=&\frac{1}{4}+\frac{1}{8}(k-1\mod 2),\\
\langle D|\hat{\mathcal{D}} ^{xy}_{k}\hat{\mathcal{D}}^{xy}_{k+\Delta}|D\rangle&=&-(-1)^{\Delta\mod 2}\frac{1}{4}.
\end{eqnarray}

This follows 
\begin{equation}
\mathcal{D}_{xy}^{\Delta}=\frac{1}{M-2-|\Delta|}\sum_{k}\langle \hat{\mathcal{D}}^{xy}_{k}\hat{\mathcal{D}}^{xy}_{k+\Delta}\rangle=\begin{cases}
-\frac{3}{8},\quad \mathrm{if} \quad \Delta=0,\\
\frac{5M-14}{16(M-3)},\quad \mathrm{if} \quad |\Delta|=1,\\
-\frac{1}{4}(-1)^{|\Delta|\mod 2}, \quad \mathrm{else}.
\end{cases}
\end{equation}
Hence, we finally get 
\begin{equation}
\begin{split}
\bar{\mathcal{D}}_{xy}&=\frac{1}{2M-5}\sum_{\Delta=-(M-3)}^{M-3}D^{xy}_{\Delta},\\
&=\frac{1}{8(M-3)}.
\end{split}
\end{equation}

For the dimer $zz$ correlator we can proceed as before and obtain 


\begin{equation}
\mathcal{D}_{zz}^{\Delta}=\begin{cases}
\frac{1}{8},\quad \mathrm{if} \quad \Delta=0,\\
-\frac{3M-8}{32(M-3)},\quad \mathrm{if} \quad |\Delta|=1,\\
\frac{1}{16}(-1)^{|\Delta|\mod 2}, \quad \mathrm{else}.
\end{cases}
\end{equation}
This implies, 
\begin{equation}
\begin{split}
\bar{\mathcal{D}}_{zz}&=\frac{1}{2M-5}\sum_{\Delta=-(M-3)}^{M-3}D^{zz}_{\Delta},\\
&=\frac{1}{(2M-5)(M-3)}\left(\frac{2-M}{16}\right).
\end{split}
\end{equation}
\vspace{4cm}
\end{widetext}

\end{document}